\documentclass[aps,pra,tightenlines,floatfix,12pt]{revtex4-1}
\usepackage{amsmath}
\usepackage{amssymb}
\usepackage{graphicx}
\usepackage{esint}
\usepackage[english]{babel}
\usepackage{bm}

\makeatletter
\@ifundefined{textcolor}{}
{%
 \definecolor{BLACK}{gray}{0}
 \definecolor{WHITE}{gray}{1}
 \definecolor{RED}{rgb}{1,0,0}
 \definecolor{GREEN}{rgb}{0,1,0}
 \definecolor{BLUE}{rgb}{0,0,1}
 \definecolor{CYAN}{cmyk}{1,0,0,0}
 \definecolor{MAGENTA}{cmyk}{0,1,0,0}
 \definecolor{YELLOW}{cmyk}{0,0,1,0}
}

\newcommand{\qql}{{\textquotedblleft}}
\newcommand{\qqr}{{\textquotedblright}\;}
\newcommand{\vc}[1]{\bm{\mathrm{#1}}}

\renewcommand{\[}{\begin{equation}}
\renewcommand{\]}{\end{equation}}

\usepackage{times}

\makeatother

\begin{document}
\global\long\def\avg#1{\langle#1\rangle}

\global\long\def\p{\prime}

\global\long\def\ket#1{|#1\rangle}

\global\long\def\bra#1{\langle#1|}

\global\long\def\proj#1#2{|#1\rangle\langle#2|}

\global\long\def\inner#1#2{\langle#1|#2\rangle}

\global\long\def\tr{\mathrm{tr}}

\global\long\def\dg{\dagger}

\global\long\def\im{\imath}

\global\long\def\pd#1#2{\frac{\partial#1}{\partial#2}}

\global\long\def\spd#1#2{\frac{\partial^{2}#1}{\partial#2^{2}}}

\global\long\def\der#1#2{\frac{d#1}{d#2}}

\renewcommand{\thefootnote}{\fnsymbol{footnote}}
\setcounter{footnote}{1}

\title{Quantum transport in ultracold atoms}
\author{Chih-Chun Chien}
\email{cchien5@ucmerced.edu}
\affiliation{School of Natural Sciences, University of California, Merced, CA 95343, USA}
\author{Sebastiano Peotta}
\email{speotta@physics.ucsd.edu}
\affiliation{Department of Physics, University of California, San Diego, California 92093, USA}
\affiliation{COMP Center of Excellence, Department of Applied Physics, Aalto University School of Science, FI-00076 Aalto, Finland}
\author{Massimiliano Di Ventra}
\email{diventra@physics.ucsd.edu}
\affiliation{Department of Physics, University of California, San Diego, California 92093, USA}

\date{\today}
\begin{abstract}
Ultracold atoms confined by engineered magnetic or optical potentials are ideal systems for studying phenomena otherwise difficult to realize or probe in the solid state because their atomic interaction strength, number of species, density, and geometry can be independently controlled.  
This review focuses on quantum transport phenomena in atomic gases that mirror and oftentimes either better elucidate or show fundamental differences with those observed in mesoscopic and nanoscopic systems. We discuss significant progress in performing 
transport experiments in atomic gases, contrast similarities and differences between transport in cold atoms and in condensed matter systems, and survey inspiring theoretical 
predictions that are difficult to verify in conventional setups. These results further demonstrate the versatility offered by atomic systems in the study of nonequilibrium phenomena and their promise for novel applications. 
\end{abstract}
\maketitle

\section{Introduction}

Transport of electrons is the backbone of modern technology since it is an extremely fast and convenient way to convey and manipulate energy and information. 
Quantum mechanics explains, by means of band theory, the astonishing variability of electrical resistivity in Nature, ranging from $10^{-8}\,\Omega\,\mathrm{m}$ in metals up to $10^{16}\,\Omega\,\mathrm{m}$ in insulators, an extremely important fact for applications. Yet this is an interesting effect by itself since, superficially, it seems to suggest that the Coulomb interaction plays a relatively little role in transport, a fact formalized within the framework of the Landau-Fermi liquid theory, which states that the charge carriers form a Fermi sea of weakly-interacting quasiparticles \cite{AshcroftBook}. 
It is instead clear that both the quantum nature of the carriers and the Coulomb interaction or other forms of interaction can have interesting and potentially useful consequences \cite{DiVentra_book,Nazarov_book}.

Take superconductivity as an example. The electron-phonon interaction in conventional superconductors gives rise to dissipationless electronic transport. Whereas conventional superconductors are consistently described by the Bardeen-Cooper-Schrieffer (BCS) theory~\cite{Schrieffer_book} which implicitly relies on the Landau-Fermi liquid paradigm, the most interesting superconductors with high critical temperature (high-$T_c$ superconductors) show strong indications of a breakdown of the quasiparticle picture and a quantitative theory thereof is still lacking~\cite{Taillefer:2010,Basov2011}. Another case where Coulomb interactions manifest most dramatically, and lead to a clear example of a breakdown of the Landau-Fermi liquid picture, is the fractional quantum Hall effect~\cite{Stone_book}, where a two-dimensional electron gas behaves as a gas of quasiparticles with fractional charge and the transport of currents occurs only at the edges.  Superconductivity and quantum Hall effect show how transport is indeed one of the most powerful probes at our disposal to investigate matter in its various phases. They also show how our current understanding of the properties of matter, in particular its transport properties, is heavily based on models where the interactions are treated approximately and the limits of validity of such approximations are often unclear.

A novel route to probe and manipulate transport of quantum matter was discovered when extremely cold (down to few nanoKelvins) and rarefied atomic gases (with typical density $n \sim 10^{13}-10^{15}\,\mathrm{cm}^{-3}$ compared to $n \sim 10^{22}\mathrm{cm}^{-3}$ in solid state systems \cite{PethickBook}) were used to realize the condensate of bosons predicted by Bose and Einstein in 1924~\cite{Davis1995,Anderson1995,Bradley1995}. These so-called ultracold gases are especially clean and controllable: as an example the interparticle interaction takes a very simple form, namely that of an isotropic contact interaction parametrized only by the two-body scattering length, which can be externally tuned \cite{PethickBook,Chin2010}, thereby accessing the entire range {\it from the noninteracting limit to the strongly interacting one}.
Elegant experiments on Bose-Einstein condensates allow detailed analyses of interacting bosons both in the ground state and out of equilibrium. The agreement with theory is quantitatively excellent meaning that this is one of the few examples of a many-body quantum system where the effect of interactions is well understood. Shortly after, the even more challenging task of cooling Fermi gases below the degeneracy temperature was accomplished~\cite{DeMarco1999}, and noninteracting Fermi gases were realized~\cite{Truscott2001}. Therefore, noninteracting bosons and fermions are no longer idealized concepts and become experimentally accessible in ultracold atoms.

These achievements marked the beginning of a new field of research  and ultracold atoms---as these extremely cold and rarefied atomic gases are called---are so controllable and flexible that they can be considered as \qql quantum simulators\qqr~\cite{Qsim1,Qsim2,Qsim3,Qsim4}, namely physical systems that can mimic the behavior of other quantum states of matter whose understanding defies conventional means. Ultracold atoms have already gained a reputation in this respect, and they are expected to tackle important unsolved problems in the future, but in the meantime the field needs very specific short term goals~\cite{Qsim1}.

\begin{figure}[t]
\includegraphics[width=3.2in]{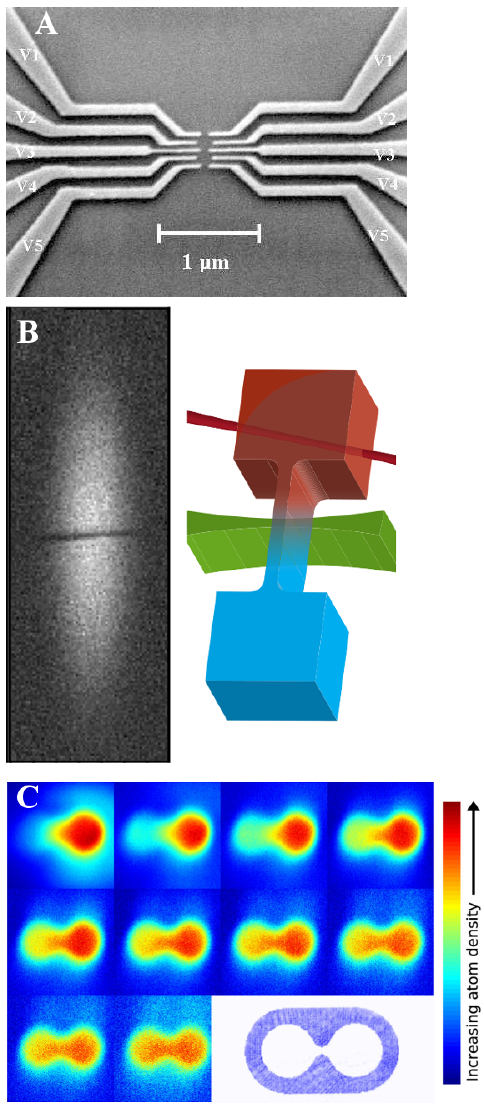}
\caption{\label{fig:openvsclosed} Contrast of transport experiments in solid-state (A) and cold-atom (B,C) systems. The quantum dot molecule illustrated in (A) is defined by gating leads and is connected to the bulk of the two-dimensional electron gas (in the upper and lower part of the figure) by tunneling barriers that can be externally tuned (from Ref.~\cite{Jeong01}). Due to the coupling between the dots and the bulk electron gas, this is an open system. In contrast, ultracold atomic systems are usually closed systems because particles only move around inside the system. In (B) one may consider the two sides of the junction as reservoirs, but the overall particle number is conserved (from Ref.~\cite{Brantut13}). In (C) the atoms flow from one side to the other in the ultracold analog of a RC-circuit implemented by confining the atoms in the dumbbell shaped region shown in the lower right (from Ref.~\cite{Lee2013}). }
\end{figure}

An ambitious yet realistic direction is to use ultracold atomic gases to study transport properties of quantum systems which do not fit into conventional descriptions. For example, one major difference between solid state systems  and ultracold gases is that the latter are usually closed systems unaffected by external environment (see Fig.~\ref{fig:openvsclosed} (b) and (c)), at least on the typical time scales  of the dynamics ranging from milliseconds to seconds. 
On the other hand, solid state systems, as the one shown in Fig.~\ref{fig:openvsclosed} (a), are intrinsically open systems and one cannot easily prevent the exchange of energy or particles with the external environment. Indeed, when a certain transport coefficient is measured, the system is assumed to be in a steady state, which is maintained by the continuous exchange of energy or particles with external reservoirs.  
It is therefore crucial to examine if  transport coefficients measured in ultracold-atom experiments can be directly compared to their counterparts that characterize transport in conventional solid state devices. This may be the case as in recent experiments that have laid the foundations for a systematic investigation of transport properties of ultracold atomic gases~\cite{Strohmaier2007,Schneider12,Vidmar2013,Ronzheimer2013,Eckel14,Salger09,Brantut12,Krinner14,
Gaunt2013,Gotlibovych2014,Schmidutz2014,Sommer11,Brantut13,Hazlett13,Cheneau12,Krinner2013,
Stadler2012,Atala2014}. The isolated nature of cold-atom systems may well turn out to be one of their greatest advantages because one can focus solely on their intrinsic transport properties, while highlighting the differences with condensed matter systems whose coupling to the environment and disorder cannot be neglected or easily engineered. 



The aim of this review is to provide an overview of the advance towards a deeper understanding of transport in quantum systems with the assistance of ultracold atomic gases experiments. In Section~\ref{sec:comparison} we introduce the reader to the methods used to control and observe these systems with emphasis on the aspects relevant to their transport properties. The main differences with analogous experimental setups in solid state physics are outlined. 
Section~\ref{sec:weak_interactions} concentrates on those phenomena that have been experimentally probed in ultracold gases which can be understood with tractable theories of weakly-interacting particles. We will discuss how these phenomena relate to their solid state counterpart while highlighting the novel features brought forth by the specifics of ultracold atoms. In Section~\ref{sec:strong_interactions} we discuss instead those experiments that have already demonstrated the potential of ultracold gases as quantum simulators by uncovering transport phenomena that are already close or beyond the limit of what can be analyzed within current analytical and numerical methods. The number of such experiments is limited, given that the idea of using ultracold gases as quantum simulators is relatively new. Nevertheless, a rapid increase in their number and sophistication is in sight.

\section{Comparison between ultracold gases and solid state systems}\label{sec:comparison}
When comparing transport phenomena in ultracold atomic gases to similar ones in conventional electronics, their differences and similarities, summarized in Table~\ref{Tab:comparison}, should be taken into account. Their importance will be appreciated when various related topics are discussed later on.
The single most important difference from our point of view is that, while electronic systems are connected to the external environment, ultracold gases are necessarily very well isolated from it due to the confinement and extremely low temperature and density (see Fig.~\ref{fig:openvsclosed}). Small external perturbations can drive an ultracold gas initially at rest out of equilibrium, often quite far from it. This is a challenge for conventional approaches to nonequilibrium dynamics based on linear response theory~\cite{MahanBook,DiVentra_book,Nazarov_book}, but at the same time an opportunity for studying transport in regimes beyond linear response that were previously inaccessible~\cite{CAdyn1,CAdyn2,CAdyn3,CAdyn4}. In any case, cold-atom systems still allow (and have been employed) to test conventional linear-response theory~\cite{Brantut13,Hazlett13}. An extreme case is the preparation of highly-excited metastable states with ``negative temperature'' relative to the motional degrees of freedom~\cite{Braun13}, a result that reignited the debate on the meaning of negative temperatures~\cite{Dunkel14} and on the consistent thermodynamic descriptions of isolated systems.

\begin{table}
\begin{tabular}{|l|p{2.2in}|p{2.2in}|}
\hline
 & Cold atoms & Electronics \\
\hline
Temperature & $10^{-9}-10^{-6}\,\mathrm{K}$ & $10^{-6}-10^{2}\,\mathrm{K}$ \\
\hline
Confinement & Optical or magnetic potentials, usually harmonic & Positively-charged ionic lattice\\
\hline
Density     &    $10^{13}-10^{15}$ cm$^{-3}$  \cite{PethickBook}  &  $\sim 10^{22}$ cm$^{-3}$ \cite{AshcroftBook}          \\
\hline
Number of components & Variable, depending on atom species & Two (spin-$1/2$) \\
\hline
Particle statistics & Bosonic or fermionic & Fermionic (electrons)\\
\hline
Natural setup    & Isolated from the environment       & Open   \\
\hline
Conventional approach & Bottom-up   & Top-down \\
\hline
Lattice geometry   & Highly tunable  & Controlled by chemical composition, pressure, etc. \\
\hline
Thermodynamic limit & $N$ and $N_p$ can be tuned separately \cite{ChienEPL12}. & $N_p/N=$ constant, $N\rightarrow\infty$. \\
\hline
Interaction & Tunable. The noninteracting limit can be accessed. & Coulomb interactions \\
\hline
Tunneling time scale & $\sim 10^{-3}$ s \cite{Ho07} & $\sim 10^{-15}$ s or less \cite{DiVentra_book} \\
\hline
Disorder & Virtually none but can be engineered & Usually unavoidable \\
\hline
Driving force & Gravity, magnetic field gradients, artificial gauge fields, local heating
& Electromagnetic field, temperature gradient, etc. \\
\hline
Main observable & Density profile & Charge current \\
\hline
\end{tabular}
\caption{\label{Tab:comparison} Summary of the differences and similarities between cold atoms and conventional electronic systems. Here $N$ is the total number of lattice sites and $N_p$ is the particle number.}
\end{table}

\begin{figure}[t]
\includegraphics[width=3.2in]{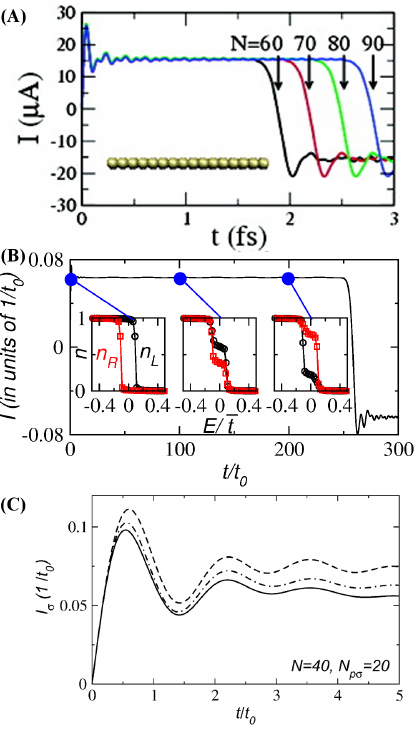}
\caption{\label{fig:QSSC} (a) A quasi-steady state current (QSSC) can emerge in an isolated quantum system of noninteracting fermions \cite{Bushong05}. (b) In an isolated system even when a QSSC is present, the particle distributions $n_L$ and $n_R$ of the two ``finite reservoirs'' evolve in time~\cite{Chien14}. (c) In the presence of onsite interactions, a QSSC still survives (from top to bottom: mean-field, mean-field plus higher-order corrections, and density matrix renormalization group results) \cite{ChienNJP13}.}
\end{figure}

A quasi-steady state, where the average current or other observables remain constant within a finite period of time \cite{DiVentra04,DiVentra_book},
is usually a necessary assumption in any theory of transport. However, far from the linear response regime it is not obvious that in an isolated system a quasi-steady state always forms and what conditions are required. It has been shown that this is the typical behavior in the case of fermions (see Figure~\ref{fig:QSSC}) , but in other systems, oscillatory behavior or more complicated dynamics can emerge instead of a quasi-steady state~\cite{Chern14,Chien14}. The Landauer theory of transport~\cite{Landauer57,DiVentra_book,Chien14} used to model transport in mesoscopic and nanoscopic systems such as the one shown in Fig.~\ref{fig:openvsclosed} (a) assumes both the existence of a steady state and that the large particle and energy reservoirs connected to the conduction channel are in the thermal equilibrium state described by the Fermi-Dirac distribution.

In ultracold atom systems the \qql reservoirs\qqr  have a much smaller size, as illustrated in Figure~\ref{fig:openvsclosed} (b) and (c), and common assumptions in conventional theories of transport~\cite{DiVentra_book,Chien14} may not necessarily hold. 
In an isolated finite system, the particle distribution usually evolves with time when a current is flowing, and a stationary Fermi-Dirac distribution may not be realistic, especially if the dynamics is driven in a particle-number conserving fashion as in ultracold gases \cite{Poli01,Ott2004,Strohmaier2007,Lin2011b,Sommer11,Eckel14,Schneider12,ChienNJP13}. An example of the dynamical evolution of particle distributions in the reservoirs is shown in Figure~\ref{fig:QSSC} (b) where an interesting observation is that the initial particle number (chemical potential) imbalance can reverse while the quasi-steady state current is still flowing in the same direction. A \qql microcanonical\qqr  approach to transport~\cite{DiVentra04,Bushong05,DiVentra_book,Chien12,Chien14}, which does not rely on the artificial distinction between the environment and system of interest (or, in other words, between a power source and its load), appears to be better suited to investigate transport in cold atoms and has already proved its usefulness as a theoretical tool~\cite{Heidrich-Meisner2010,Chien12,Bruderer2012,Gallego-Marcos2014,Chien14}.

Indeed the microcanonical approach 
serves as a bridge between the conventional transport setup, where distinct power source and load are present, and isolated systems where the particle number is conserved exactly. It is also very well suited to the study of a quantum quench, namely a sudden change of a parameter used to excite a response, which is the most common kind of experiment performed with ultracold gases~\cite{CAdyn2}. Transport phenomena that are currently analyzed using conventional linear response theory~\cite{Brantut12,Stadler2012,Krinner2013,Brantut13,Hazlett13,Krinner14} may provide further information when fully microcanonical analyses are implemented. On the other hand, cold-atoms systems where atoms are continuously loaded in a trap and removed with a terminator beam, i.e., the cold-atom analogue of a battery, have been realized~\cite{Zozulya13}. In this case, appropriate modifications to the microcanonical formalism should be implemented~\cite{Chern14}.  

Dynamics out of equilibrium can also invalidate the quasiparticle description, which is appropriate only if both the quasiparticle excitation energy and the number of excited quasiparticles are small. This is because quasiparticle properties are well-defined only if the system is close to the ground state, while real particles are true eigenstates of the Hamiltonian with infinite lifetime. In this respect, very useful lessons can be learned from ultracold gases. This new trend in the study of nonequilibrium many-body phenomena using cold atoms \cite{CAdyn1,CAdyn2,CAdyn3,CAdyn4} will be discussed in Section~\ref{sec:strong_interactions}.

In cold-atom systems the absence of a natural confinement, like the positively charged background for electrons provided by the ionic crystal, has some major consequences for transport. On the one hand, the confinement of the atoms has to be engineered artificially using optical or magnetic means \cite{PethickBook,Dalibard:2011,Qsim2}. On the other hand, this allows flexibility in shaping the confining potential, a freedom that has been exploited at large thanks to the recent advancements in the techniques to tailor complex potentials for trapping and moving atoms into desired patterns~\cite{Salger09,Brantut12,Krinner14,Gaunt2013,Gotlibovych2014,Schmidutz2014,Sommer11,Brantut13,Krinner2013,Stadler2012}. The typical confinement is the harmonic trap \cite{PethickBook}, but recently it has been possible to engineer box-shaped potentials~\cite{Gaunt2013,Gotlibovych2014,Schmidutz2014} which are ideal for the study of transport phenomena in uniform setups~\cite{Peotta14_SFMI}. It is also possible to simulate \qql reservoirs\qqr to play the role of the external \qql environment\qqr in conventional solid state systems and ensure that the dynamics in the small region of interest (the \qql conducting channel\qqr or \qql load\qqr) is in a quasi-steady state for a finite period of time~\cite{Brantut12,Krinner2013,Krinner14}. Moreover, the ability to control the external confinement offers multiple options for driving cold atoms out of equilibrium, for example, by using gravity to pull atoms \cite{Poli01}, displacing the trapping potentials or loading the atoms away from their equilibrium positions \cite{Ott2004,Strohmaier2007,Sommer11,Eckel14}, creating inhomogeneous density profiles~\cite{Schneider12,Chien12} or interactions \cite{ChienNJP13}. Lately, it has also been possible to realize artificial electric and magnetic fields that can drive atoms despite their charge neutrality \cite{Lin2009a,Lin2009b,Lin2011b,Dalibard:2011,PS1:2012,PS2:2012} and bring cold gases a lot closer to solid state systems.

Optical lattices are a remarkable technique employed in cold-gases experiments, which allows realizations of analogues of the periodic ionic potential of crystals. By tuning the intensity of the laser generating the optical lattice, the lattice depth varies accordingly and the tunneling between different sites can be controlled \cite{Greiner02,PethickBook}. The interaction between atoms can also be tuned by using magnetic fields to control atomic collisions \cite{Chin2010,PethickBook}. Thus the ratio between interaction energy and kinetic energy can be increased thereby driving the atoms into a correlated regime. For bosons this eventually induces a transition from a superfluid state to a Mott insulator at integer filling of the lattice~\cite{Greiner02}. Among the lattice structures in conventional solid state systems~\cite{AshcroftBook}, the cubic, square, triangular, honeycomb, checkerboard, kagome, and many other lattices have been realized for trapping cold atoms~\cite{Greiner02,Struck11,Panahi11,Tarruell12,Jo12}. It is also possible to induce complex tunneling coefficients either by using artificial gauge fields or by modulating the lattice, which realizes the Peierls substitution for lattice systems in the presence of magnetic flux~\cite{PS1:2012,PS2:2012}. This is important in order to realize the quantum Hall effect with the associated chiral edge states~\cite{Mancini:2015,Stuhl:2015}, despite the charge neutrality of the atoms employed. Many interesting topological phases and their associated transport phenomena can be investigated using ultracold atoms in optical lattices \cite{Sato09,Goldman10,Sun12,Goldman13}. The finite size of cold-atom systems naturally provides the system boundary where interesting edge states and their associated transport phenomena take place. Another important feature of artificial lattice potentials is that they can be made species-dependent so in a multi-component system different components may see different confining geometries \cite{Mandel03,Panahi11,McKay13}.

In conventional electronic systems Coulomb interactions tie the electrons and their ionic backgrounds together and any imbalance relaxes very quickly to restore charge neutrality. For semiconductors this happens on a timescale $t \sim 10^{-11}\,\mathrm{s}$~\cite{Grosso_book}; for metals on an even shorter time. In contrast, in cold-atom systems the particle density has in general large variations which can be directly measured by absorption imaging~\cite{PethickBook} and this is usually the main observable in cold atom experiments. By preparing an ensemble of systems with identical initial condition and time evolution, one can take snapshots at different times and obtain a series of density profiles. The current is then related to the density profile~\cite{Hung2010,Schneider12,Chien12} by the continuity equation. While a more faithful atomic analogue of the STM has been proposed \cite{KollathSTM}, imaging and manipulations at the single-atom level using optical methods have been demonstrated~\cite{Buecker2009,Bakr2009,Sherson2010}. 


Another feature of ultracold gases is the intrinsic bottom-up approach where building blocks such as various species of atoms and lattice geometries can be assembled on demand~\cite{Strohmaier2007,
Schneider12,Vidmar2013,Ronzheimer2013,Eckel14,Salger09,Brantut12,Krinner14,Gaunt2013,Gotlibovych2014,
Schmidutz2014,Sommer11,Brantut13,Hazlett13,Cheneau12,Krinner2013,Stadler2012,Atala2014,
CAdyn2,CAdyn3,CAdyn4,Jochim:2015}. Instead, the approach in conventional condensed matter systems is usually top-down~\cite{MahanBook,DiVentra_book,Nazarov_book}, where devices are fabricated with carriers such as electrons already built in and shrinking the device size is an important goal. The bottom-up approach is particularly relevant in nanoscale systems, and lessons learned from ultracold atomic systems are valuable. 

\begin{figure}
\includegraphics[scale=0.7]{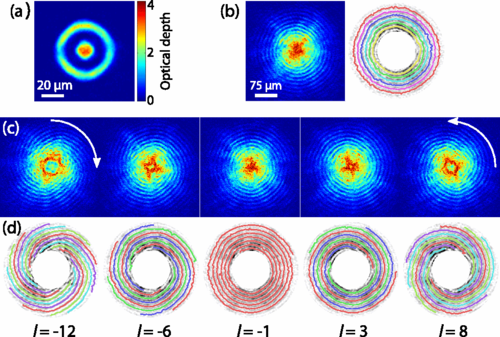}
\caption{\label{fig:interferograms} \textbf{a}) Absorption imaging of a ring-shape condensate with another condensate in the middle. The interference observed upon releasing the inner condensate can be used to measure the total current flowing in the ring. Panel \textbf{b}) shows the interference patter obtained in absence of current, while the spiralling patterns shown in \textbf{c}) and \textbf{d}) correspond to the finite amount of angular momentum quanta $l$ shown at the bottom. From Ref.~\cite{Eckel2014}.}
\end{figure}

\section{Transport in weakly-interacting gases}\label{sec:weak_interactions}
In a large number of experiments with ultracold gases it has been possible to successfully reproduce 
transport phenomena in solid state systems that have been long studied, usually within a framework of noninteracting or weakly-interacting particles. These phenomena can be sometimes challenging if not impossible to access experimentally in solid state systems and ultracold gases are an excellent opportunity to revisit them, at the same time laying the foundations for  tackling more sophisticated models and benchmarking available experimental techniques for the preparation, manipulation and the measurement of these systems. Some significant examples are Bloch oscillations that defy a clear demonstration in solid state systems but nevertheless have been elegantly realized using cold atoms in optical lattice~\cite{Dahan96,Poli01,Morsch2001}, Wannier-Stark ladder in periodic potentials~\cite{Wilkinson1996,Morsch2006}, and Landau-Zener tunneling~\cite{Morsch2001,Morsch2006}. These demonstrations have been possible due to the absence of a significant amount of disorder, an intrinsic feature of cold atoms. Moreover, adding a finite collision rate to the carriers allows the observation of negative differential conductance~\cite{Ott2004,Labouvie14}, which is found to be reasonably well-modeled by the classical Esaki and Tsu model~\cite{NDC1,NDC2} for semiconductor superlattices. Negative differential conductance has been predicted to occur also in systems with spatially dependent interactions~\cite{ChienNJP13}, which can be realized in ultracold gases.

A large body of work has been devoted to the study of the properties of weakly interacting Bose gases, for which the Gross-Pitaevskii equation has emerged as a simple description of gases deep in the condensed phase, namely well below the critical temperature for Bose-Einstein condensation~\cite{Dalfovo1999,PethickBook}. This equation describes the dynamics of the wavefunction occupied by a macroscopic number of particles with the interactions accounted for by a nonlinear term. These systems are clear  manifestations of the wave nature of matter and are manageable both theoretically and experimentally. Excellent reviews of the subject are available~\cite{Dalfovo1999,PethickBook,Qsim2} and therefore we will not discuss the topic any further. 
We only mention that recently a box-like confining potential has been realized and used to observe the properties of a uniform condensate and the Joule-Thompson effect~\cite{Gaunt2013,Gotlibovych2014,Schmidutz2014}. Interesting applications of this technique for transport-related studies of ultracold gases are expected to provide a direct comparison with those in solid state systems (see for example~\cite{Peotta14_SFMI}), while the conventional harmonic confinement of cold atoms introduces inhomogeneity that complicates the theory and the interpretation of data. 

Ultracold gases are also an excellent laboratory for the study of superfluid properties. Both bosonic and fermionic superfluids are available and the range from weak to strong interactions can be accessed \cite{Chin2010,PethickBook,Leggett}. A smoking gun of superfluid transport is the quantization of vorticity and in fact direct imaging of arrays of vortex cores in atomic Bose-Einstein condensates have been observed and provided an unambiguous proof of superfluidity in ultracold gases~\cite{Abo-Shaeer2001}. A persistent current in a ring geometry is an equilibrium property (Ref.~\cite{Nazarov_book} and references therein) which has been found in superconductors \cite{Deaver61} and superfluid helium \cite{Kojima72} by applying a magnetic flux through the ring and by rotation. Advances in shaping the laser-induced potential have made it possible to generate atomic persistent currents in single-component bosons \cite{Ryu07,Wright13} (see Fig.~\ref{fig:interferograms}) as well as in spinor bosons \cite{Beattie13} and to observe in real time the phase-slip processes that lead to the decay of the current. Theoretical analyses show that the current is expected to survive in a broad range of interactions and inhomogeneity \cite{Song09,Cominotti14}. Hysteresis of the quantized persistent currents has been reported~\cite{Eckel14}, an effect that has been elusive in both the much studied superfluid helium and superconductors. Moreover, an optical potential barrier inserted in the ring plays the role of a Josephson junction~\cite{Levy2007}, whose current-phase relation can be measured, e.g, by interfering the ring-shaped condensate with another condensate located in the middle of the ring~\cite{Eckel2014} (see Fig.~\ref{fig:interferograms}).  
Inserting two such junctions/barriers allows the realization of the ultracold gas analogue of a SQUID~\cite{Ryu13}, a device with interesting applications as a high-precision rotation sensor. The superfluid fraction, which should be distinguished from the condensate fraction \cite{Leggett}, of ultracold atomic gases can be measured by means of the second sound~\cite{Sidorenkov2013} or using an analogue of the classical Andronikashvili experiment~\cite{Cooper2010}.

The toolbox of cold-atom research has been recently enriched by techniques to realize artificial gauge fields~\cite{Lin2009a,Lin2009b,Lin2011a} and artificial spin-orbit coupling~\cite{Lin2011b,Galitski13}, which add a whole new dimension to the study of transport in ultracold gases. Artificial gauge fields are obtained using the Berry phase induced by dressing atomic states with laser fields, thereby sidestepping the limitation of charge neutrality of cold atoms. The effect of an artificial magnetic field on a Bose-Einstein condensate has been recently measured and interpreted as a form of superfluid Hall effect~\cite{LeBlan2012}. By using a uniform nonabelian  gauge field, which is equivalent to a spin-orbit coupling term, the bosonic analogue of the spin Hall effect has been detected~\cite{Beeler2013}. Several techniques have also been used to introduce artificial magnetic fluxes in lattices in the form of the usual Peierls substitution using either Raman-assisted hopping~\cite{PS1:2012, Aidelsburger:2013, Miyake:2013,Atala2014}, lattice shaking~\cite{Struck11,PS2:2012, Jotzu:2014} and synthetic dimensions~\cite{Mancini:2015,Stuhl:2015}. 

In the works described in Ref.~\cite{Aidelsburger:2013} and~\cite{Miyake:2013} Raman-assisted hopping has been used to realize the Harper Hamiltonian~\cite{Harper:1955} consisting in the combination of a large uniform (pseudo-)magnetic flux with a lattice potential and characterized by the fractal behavior of the spectrum as a function of magnetic field~\cite{Hofstadter:1976}. Large (pseudo-)magnetic field means that the flux piercing a primitive cell in the lattice is a sizable fraction of the (pseudo-)flux quantum. In these works the fractal spectrum was not observed, but the dynamics of the atoms can be visualized in real time and is consistent with the presence of a large (pseudo-)magnetic field. Another interesting feature of Raman-assisted hopping is that the artificial magnetic field can have opposite signs for different spin components~\cite{Aidelsburger:2013}, which means that time-reversal invariance is preserved, an essential ingredient for a cold-atom implementation of the recently discovered quantum spin Hall effect~\cite{Konig:2007}. A more detailed analysis of various transport phenomena and the use of fermionic species instead of bosonic $^{87}$Rb currently employed will be among the next major goals.

The technique of lattice shaking generally leads to breaking of time-reversal symmetry~\cite{Struck11,PS2:2012}, but on the other hand this latter method is simple to realize in practice and quite flexible. One of its greatest success has been the implementation of the Haldane model~\cite{Jotzu:2014}. The Haldane model was deemed impossible to realize in the solid state by its inventor~\cite{Haldane:1986} since its main ingredient includes next nearest-neighbor complex hopping coefficients in a hexagonal (graphene-like) lattice which correspond to a nontrivial magnetic field with zero average in the primitive cell. Jotzu \textit{et al.}~\cite{Jotzu:2014} were inspired by a previous work~\cite{Oka:2009} predicting that a Hall conductance could be generated in graphene when irradiated with circularly polarized light. It turns out that elliptical lattice shaking (out-of-phase shaking in the $x$ and $y$ directions) has an effect somewhat similar to circularly polarized light, that is introducing sizable complex hoppings that gap out the Dirac cones in the band structure of the hexagonal lattice. The two resulting bands, well-separated in energy, are characterized by a nontrivial topological invariant, the Chern number, an integer given by $\nu = \frac{1}{2\pi} \int_{B.Z.} d^2\vc{k}\,\Omega(\vc{k})$ where the integral is extended over the whole two-dimensional Brillouin zone and the Berry curvature $\Omega(\vc{k})$ is defined as  
$\Omega(\vc{k}) = i\left(\left\langle \frac{\partial u_{\vc{k}}}{\partial k_x}\bigg|\frac{\partial u_{\vc{k}}}{\partial k_y} \right\rangle - \left\langle\frac{\partial u_{\vc{k}}}{\partial k_y}\bigg|\frac{\partial u_{\vc{k}}}{\partial k_x} \right\rangle \right)$
in terms of the periodic Bloch functions $u_{\vc{k}}(\vc{r})$. The Chern number characterizes the Hall conductance in a completely filled band, which is given by $\sigma_{xy} = \frac{e^2}{h}\nu$, while the Berry curvature acts as an effective magnetic field in the semiclassical equations of motion that govern the dynamics of a wavepacket in a Bloch band. Wavepacket dynamics is the main tool used for measuring the Berry curvature in the experimental realization of the Haldane model~\cite{Jotzu:2014} thereby providing access to the topological properties of the band structure. Interestingly, the Chern number of the Harper model has been measured in cold atoms~\cite{Aidelsburger15}, too.

Another possibility for the generation of a large (pseudo-)magnetic field is that of synthetic dimensions~\cite{Celi:2014}, which has been recently implemented~\cite{Mancini:2015,Stuhl:2015}. 
The idea is to use the internal states of neutral atoms as a fictitious spatial dimension. The advantage of using such a mapping from internal states to orbital states is that it is much easier to introduce complex hopping in this synthetic dimension than in a real dimension by means of the Raman-induced transition between the internal states. Moreover, since the number of internal states is finite, this allows to realize an essentially perfect ribbon geometry, which is ideally suited for the observation of chiral edge states induced by a (pseudo-)magnetic field. The dynamics of the chiral edge states can be visualized by means of selective imaging of the atom internal states, which corresponds to a position measurement along the fictitious dimension. Such well-defined geometries and high-resolution spatial imaging of cold atoms are expected to reveal more exciting topological aspects of transport.

The available options for introducing artificial magnetic fields and other exotic gauge fields are numerous, with some proposals still awaiting to be realized. Each technique has its own advantages and drawbacks. For instance, an important issue is the heating of the atoms induced by Raman transitions or lattice shaking. For studying transport phenomena, time-dependent artificial gauge fields can also be used to induce quasi-steady state currents in ultracold fermions and bosons~\cite{ChienPRA13,Peotta14_SFMI} and this includes the possibility for engineering a time-dependent spin-orbit interaction, something which is unusual in condensed matter systems. 

\begin{figure}
\includegraphics[scale=0.25]{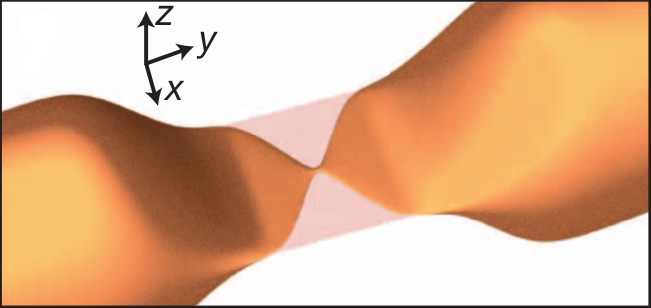}
\includegraphics[scale=0.4]{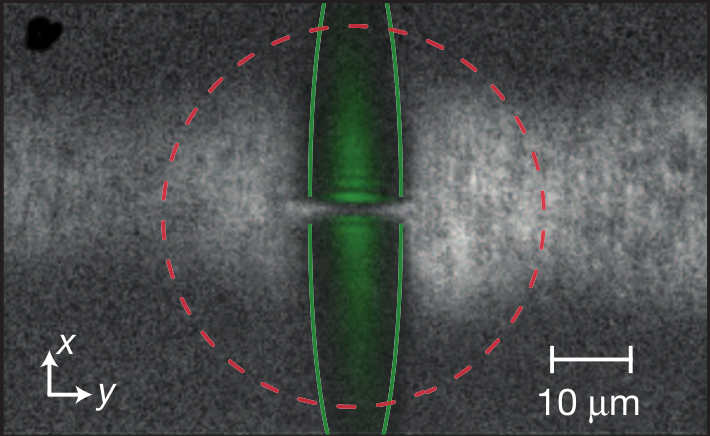}
\caption{\label{fig:quantized-conductance} (left) Schematic three-dimensional view of the atomic quantum point contact used to measure the phenomenon of quantized conductance in Ref.~\cite{Krinner14}. 
A laser in the $\text{TEM}_{01}$ mode is used to produce a thin conducting layer in an elongated atomic trap for fermionic $^{6}\text{Li}$ atoms. 
A constriction along the $x$ direction is lithographically imprinted by another laser beam. The resulting channel supports few transverse modes that can be selectively populated.
(right) False color image of the atomic density (green). The red dashed line indicates the intensity profile of the laser beam used to tune the chemical potential and thus the transverse mode population in the channel.}
\end{figure}

Recently, in a set of important experiments by the Esslinger's group at ETH~\cite{Brantut12,Krinner14,Brantut13,Krinner2013,Stadler2012} it has been possible to realize and investigate a cold-atom setup that mimics very well the minimal model used to describe conduction in mesoscopic systems. In the latter case it is usually assumed that two ideally infinite reservoirs of particles are connected by means of a constriction (channel). One of the most important results of the theory of transport in mesoscopic systems is that even if the channel is ballistic, the conductance of the overall system, including the large reservoirs of carriers, is not diverging as one would naively expect, but has a finite value~\cite{DiVentra_book}. The origin of this finite resistance has been the subject of heated discussions at the dawn of the theory of transport in mesoscopic solid state systems and nowadays it is generally accepted to be an effect whereby all dissipative processes occur in the large particle reservoirs, whereas no relaxation processes take place in a ballistic channel. Moreover, under certain conditions, such as low temperatures and the adiabatic connection of the reservoirs to the channel in order to prevent backscattering of the incoming particles~\cite{Glazman:1988}, the conductance is quantized in units of the fundamental constant $G_0 = 2q_{e}^2/h \approx 7.75 \times 10^{-5} S$, namely each transverse mode in the ballistic channel contributes an amount $G_0$ to the total conductance. For cold atoms, the mass current is analogous to the electric current so the conductance is well-defined by setting the charge $q_{e}=1$. The experimental observation of quantized conductance~\cite{Wees:1988,Wharam:1988}, as predicted by the Landauer theory of transport, has been an important achievement for the then rising field of mesoscopic systems, and can be considered as an analogous milestone in the rising cold-atom technology.

\begin{figure}
\includegraphics[scale=2.0]{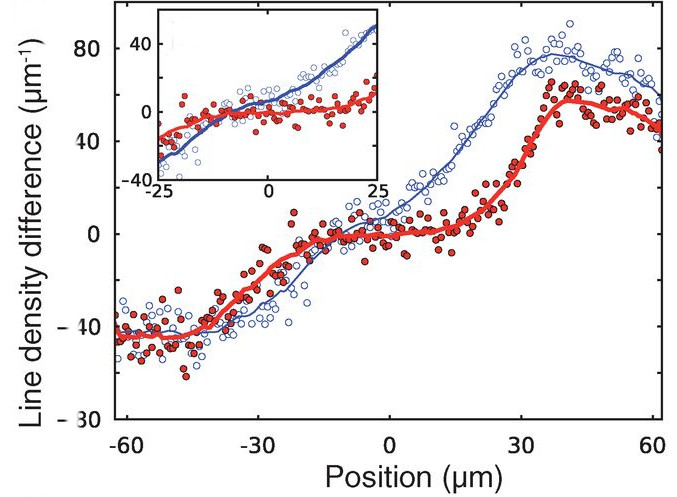}
\caption{\label{fig:density-profiles} Difference of the density profiles between a current-carrying state (finite population imbalance) and a state with zero current in the region of the channel constriction in the setup shown in Fig.~\ref{fig:quantized-conductance}. The red full dots refer to a ballistic channel while in the case of the empty blue dots disorder has been artificially introduced. From Ref.~\cite{Brantut12}.}
\end{figure}

\begin{figure}
\begin{tabular}{cc}
\includegraphics[scale=0.25]{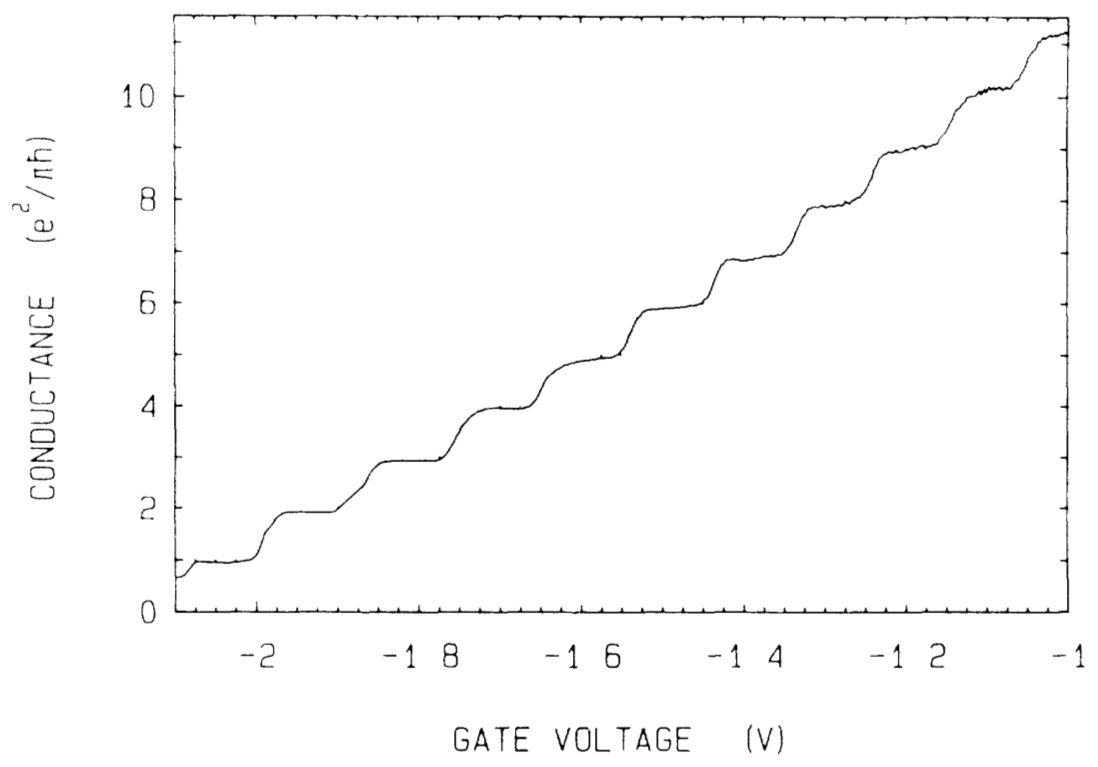} & 
\includegraphics[scale=0.2]{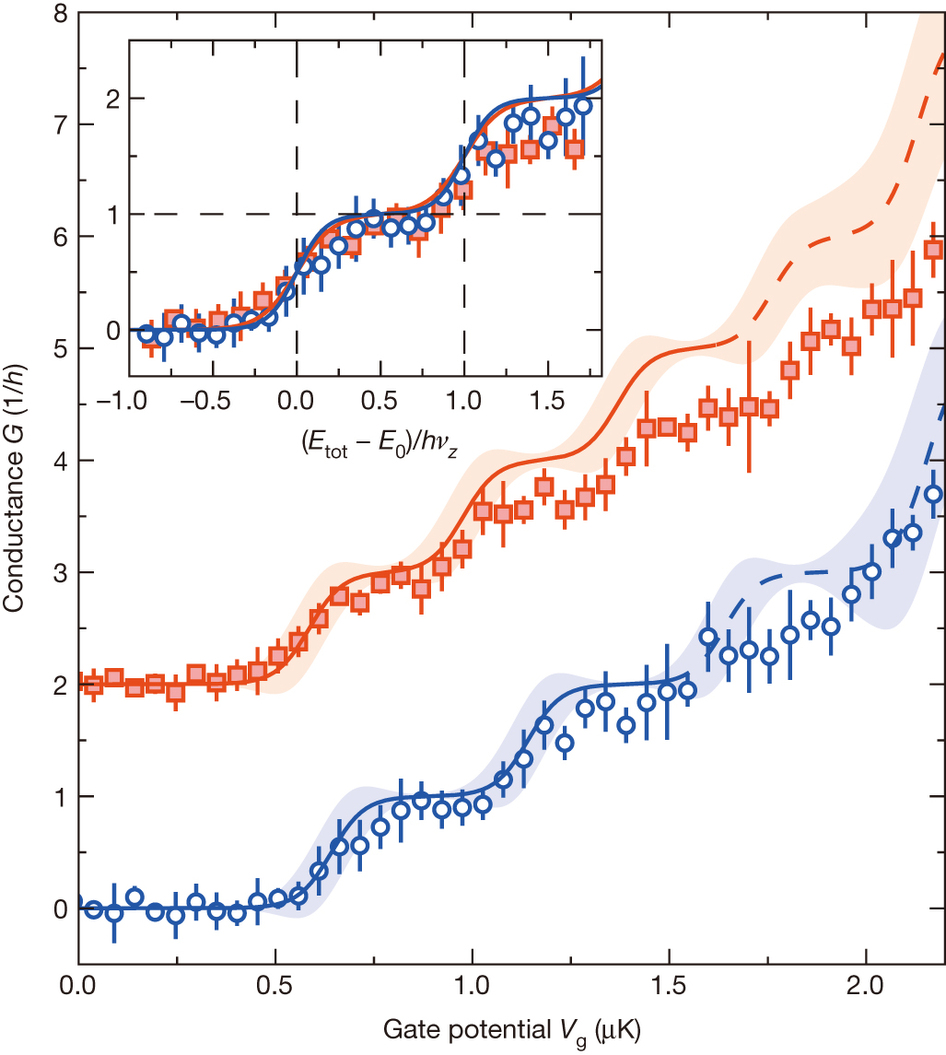}
\end{tabular}
\caption{\label{fig:quantized-conductance-2} Quantized conductance plateaus in a 2D dimensional electron gas [2DEG] (left) from Ref.~\cite{Wees:1988}, and in the cold atom setup shown in Fig.~\ref{fig:quantized-conductance} (right), from Ref.~\cite{Krinner14}. In the case of electronic transport the conductance is measured in units of $G_0 = 2e^2/h$ while the equivalent unit in the case of neutral matter is $G_{0,\text{neut.}} = 1/h$. Relevant parameters in the 2DEG: temperature $T = 0.6\,\text{K}$, constriction width in correspondence of the first plateau $W \approx 20\,\text{nm}$, Fermi wavelength $\lambda_F \approx 40\,\text{nm}$, mean free path $\ell = 8.5\,\mu\text{m}$, areal density $n = 3.56 \times 10^{11}\,\text{cm}^{-2}$. Relevant parameters in the cold atom experiment: $T = 42\,\text{nK}$, constriction width  $W \approx 1.5\,\mu\text{m}$, Fermi wavelength $\lambda_F \approx 2\,\mu\text{m}$, collision mean free path $\ell = 12\,\text{mm}$, density $n = 10^{13}\,\text{cm}^{-3}$.}
\end{figure}

In the experiments performed in Refs.~\cite{Brantut12,Krinner14,Brantut13,Krinner2013,Stadler2012}
a laser beam in the $\text{TEM}_{01}$ mode is used to squeeze the central part of an elongated harmonic trap in order to form a quasi-2D conducting channel. The channel forms at the intensity node of the laser which acts as a repulsive potential, while the left and right regions away from the constriction play the role of particle reservoirs (Fig.~\ref{fig:quantized-conductance}). 
If the system is prepared in a state with a finite population imbalance a current flow is established 
which results in the exponential decay of the particle number difference in the reservoirs analogous to the discharge of an RC circuit (see also~\cite{Lee2013} and Fig.~\ref{fig:openvsclosed} (c)). Since the effective \qql capacitance\qqr of the circuit can be reconstructed from the well-known geometry of the confinement, it is possible to infer the conductance of the channel.
A finite contact resistance can be measured even in the case of a ballistic channel, i.e., in the case of a collision-induced mean free path larger than the length of the channel~\cite{Brantut12}, in agreement with Landauer theory. The advantage of using ultracold gases is that the particle density can be recorded in real time and in the same experiment it has been verified that the particle density is constant in the case of ballistic conduction and linearly decreasing for a diffusive channel realized by artificially introducing disorder~\cite{Brantut12}, as shown in Fig.~\ref{fig:density-profiles}. By further shaping the conducting region into a quasi-1D constriction as shown in Fig.~\ref{fig:quantized-conductance} (left) one enters the regime where only few transverse modes in the channel participate in the conduction  and it is possible to resolve the quantized conductance plateaus~\cite{Krinner14}. Fig.~\ref{fig:quantized-conductance-2} shows precisely the result of this experiment performed with ultracold gases side by side with the original data obtained with a 2D electron gas in a GaAs-AlGaAs heterostructure~\cite{Wees:1988}. The agreement between the two systems is remarkable 
given the difference in the length and energy scales that characterize them, the most important ones are summarized in the figure caption. Perhaps the most surprising finding from the cold atom experiment is that, in the reservoirs, neither a short mean free path  nor sizable interactions are needed for the observation of quantized conductance. Indeed the reservoirs are essentially in the collisionless regime, the mean free path being 40 times the size of the trap, and quantized conductance has been observed even in the limit of noninteracting particles~\cite{Krinner14}. This indicates that the value of the quantized conductance (and hence the existence of a quasi-steady state) is not directly related to the existence of well-defined 
particle distributions (see, e.g., Ref.~\cite{Chien14}).

Using essentially the same set up several other phenomena have been probed such as the effect of superfluid flow~\cite{Stadler2012}, the interplay between superfluidity and disorder~\cite{Krinner2013}, and even thermoelectric transport~\cite{Brantut13} in its equivalent form for neutral matter. Interestingly, in the latter case an extremely high value, compared to current solid state standards, of the atomic analogue of the dimensionless figure of merit $ZT \approx 2.4$ has been reported. These elegant demonstrations show how ultracold gases can be used to provide insights into important open problems of condensed matter physics, such as increasing the efficiency of thermoelectric materials. The results so far agree with the predictions of the Landauer theory and the microcanonical approach to transport~\cite{DiVentra_book,DiVentra04}. An interesting direction for  the future is to study the corrections to the Landauer theory of transport due to many-body effects as outlined in Ref.~\cite{Vignale2009}.

Whereas the main focus so far has been on mass transport, we also note a number of recent studies on spin and thermal transport in one-dimensional spin chains at zero and finite temperatures~\cite{Karrasch2013a,Karrasch2013b,Karrasch2014} by means of sophisticated quasi-exact numerical methods. These problems may be mapped to systems that can be simulated by cold atoms, which will then provide well controlled tests for numerical studies of transport phenomena in many-body systems. Other interesting transport phenomena such as spin-charge separation in interacting 1D fermions \cite{Kollath05,Enss12} have been studied in numerical simulations. 

As illustrated in Figure~\ref{fig:QSSC}, in an isolated quantum system a quasi-steady state current can emerge. In this respect, several theoretical predictions have yet to be verified. For example, particle statistics is expected to play an essential role~\cite{DiVentra04} since a quasi-steady state inevitably forms for weakly-interacting or noninteracting fermions while a finite interaction is indispensable in the case of bosons~\cite{Chien12,Peotta14_SFMI}. Using quasi-exact numerical methods in one dimension it has been verified that a quasi-steady state occurs for all interaction strengths of bosons when the system is not a Mott insulator~\cite{Peotta14_SFMI} (see also  Section~\ref{sec:strong_interactions}), while a large driving bias or interactions in fermions can lead to dynamically generated insulating states \cite{ChienNJP13}. In contrast, in lattice systems where a flat band of dispersionless and localized states is present, it has been predicted that strong fluctuations undermine the formation of the quasi-steady state current~\cite{Chern14} due to the interference of localized and mobile particles. 


\begin{figure}
\includegraphics[scale=0.4]{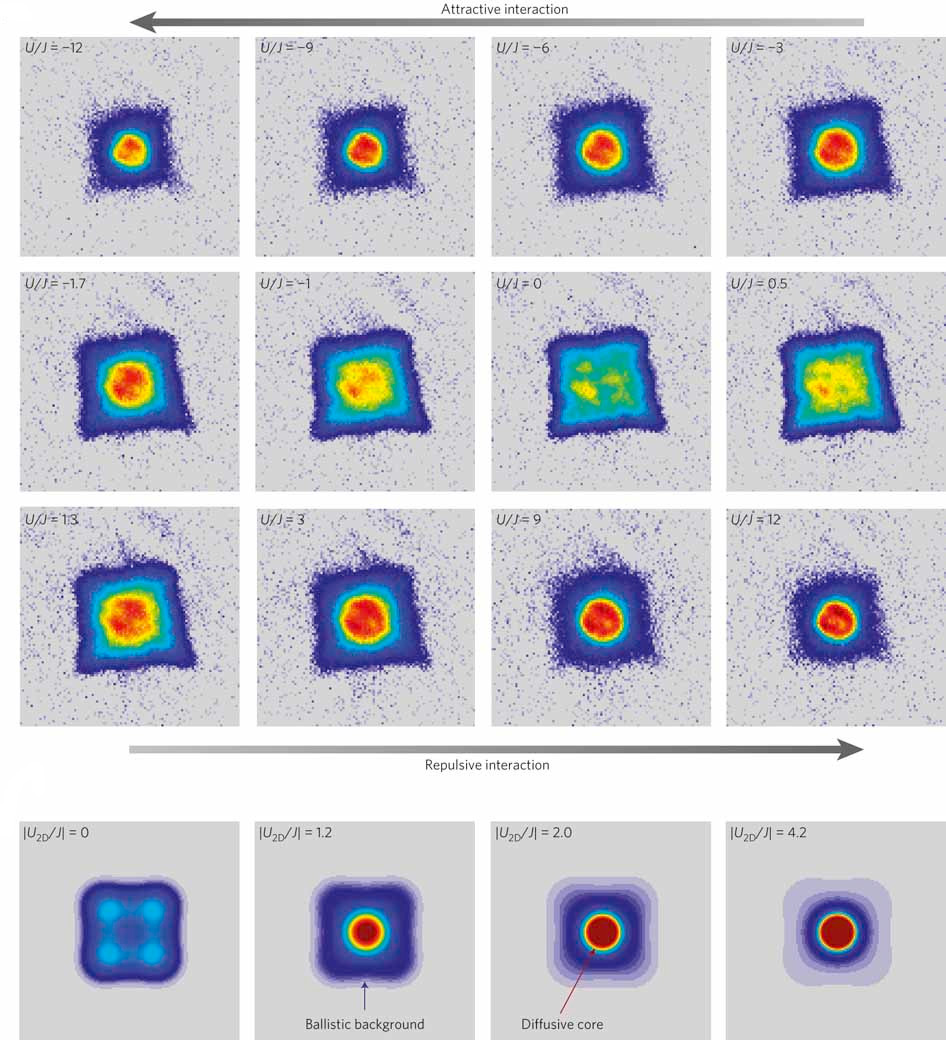}
\caption{\label{fig:expansion} Free expansion of fermionic atoms in a two dimensional optical lattice for various values of the Hubbard interaction $U/J$ from attractive to repulsive. In the noninteracting case the gas expands ballistically and the shape changes from spherically symmetric to square, which is the symmetry of the underlying lattice. Instead, in the interacting case a large portion of the gas remains spherically symmetric, an indication of relaxation to thermal equilibrium, while only the low density tails have a square shape due to the ballistic motion. An approach based on the Boltzmann equation can qualitatively capture the dynamics (bottom). Figure taken from Ref.~\cite{Schneider12}.}
\end{figure}

\section{Transport in strongly-interacting gases}\label{sec:strong_interactions}
The tunability of the interactions among atoms using external magnetic fields~\cite{Chin2010,PethickBook} and the ability to trap an atomic gas in artificial periodic potentials are two important ingredients for accessing strongly interacting states of quantum gases. By tuning the effective contact interaction, the unitary Fermi gas, where two-body bound states start to emerge (see \cite{PethickBook,Leggett,BECBCSbook} and references therein), has been realized and analyzed in great detail. Fundamental transport coefficients such as the shear viscosity of these strongly interacting systems have been measured~\cite{Cao11}. 
The shear viscosity $\eta$ at low temperatures is expected to be on the order of $\hbar$ times the density while at higher temperatures the scaling $\eta \sim T^{3/2}$ is expected, and both results have been verified~\cite{Cao11}. The ratio $\eta/s$, where the entropy density $s$ can be derived from the equation of state, is close to a conjectured universal lower bound $\eta/s\sim\hbar/k_B$~\cite{Kovtun05,Adams12}. There has been a debate on whether the unitary Fermi gas will exhibit a minimum in $\eta$ or $\eta/s$ as its temperature decreases \cite{Guo11,Enss11}, and a recent experiment~\cite{Thomas14} suggests a minimum in $\eta/s$ but not in $\eta$. 

Spin transport has also been studied extensively in the unitary Fermi gas, where the spin refers in this context to the internal states arising from the hyperfine structure of the atoms. The longitudinal $D_s^{\perp}$~\cite{Sommer11} and transverse $D_s^{\parallel}$~\cite{Bardon14} spin diffusion constants have been found consistent with the quantum limit $ D_s \gtrsim \hbar/m$ which can be deduced from dimensional analysis. In the case of the transverse spin transport, where the magnetization is perpendicular to the magnetic-field gradient, Leggett and Rice~\cite{Leggett1968,Legget1970} predicted a simultaneous precession of the spin in addition to the diffusion, an effect which has been observed in the unitary Fermi gas~\cite{Trotzky15}. In addition, spin transport in a two-dimensional Fermi gas has also been investigated~\cite{Kosch2013} and in the strongly-interacting regime the spin diffusion constant also approaches the quantum limit from above.
 
As mentioned, the ratio between the kinetic (hopping) energy and interaction energy of cold atoms in optical lattices can be tuned continuously. This can result in transitions to strongly correlated phases such as the bosonic Mott insulator demonstrated in a landmark experiment in 2002 ~\cite{Greiner02}. There have been numerous investigations of transport properties of the Bose- and Fermi-Hubbard models~\cite{Strohmaier2007,Schneider12,Cheneau12,Vidmar2013,Ronzheimer2013}, whose Hamiltonians are respectively
\begin{equation}\label{eq:bose-hubbard}
\mathcal{H}_{\text{B}} = -J\sum_{\langle ij\rangle} \left(\hat{b}^\dagger_{i}\hat{b}_{j} + \hat{b}^\dagger_{j}\hat{b}_{i}\right) + \frac{U}{2}\sum_i \hat{b}^{\dagger 2}_i \hat{b}^2_i\,,
\end{equation}
and
\begin{equation}\label{eq:fermi-hubbard}
\mathcal{H}_{\text{F}} = -J\sum_{\langle ij\rangle,\sigma} \left(\hat{c}^\dagger_{i\sigma}\hat{c}_{j\sigma} + \hat{c}^\dagger_{j\sigma}\hat{c}_{i\sigma}\right) + U\sum_i \hat{c}^{\dagger}_{i\uparrow}\hat{c}_{i\uparrow} \hat{c}^{\dagger}_{i\downarrow}\hat{c}_{i\downarrow}\,.
\end{equation}
The operators $\hat{b}_i$ and $\hat{c}_{i\sigma}$ are the annihilation operators on site $i$, for bosons and fermions, respectively. The energy $J$ parametrizes the nearest-neighbor hopping ( $\langle ij\rangle$ denote pairs of neighboring sites) and $U$ parametrizes the Hubbard onsite interaction. While bosons are allowed to self-interact, Pauli exclusion principle suppresses the main collision process between identical fermions so it takes at least two components of fermions to realize the Hubbard model. The two (or more) components may be different species of atoms or the same species of atoms in different internal quantum states. The ratio $U/J$ is controlled by the intensity of the optical lattice laser or the magnetic field controlling the atomic collisions. Free expansions of initially confined bosonic~\cite{Vidmar2013,Ronzheimer2013} and fermionic~\cite{Schneider12} atomic gases in an optical lattice has been studied experimentally and some experimental data in the case of fermionic atoms are shown in Fig.~\ref{fig:expansion}. It has been found that far from the integrable (exactly solvable) limit the system displays diffusive behavior with a bimodal expansion, namely a slowly diffusing central region and a rapidly expanding halo where transport is ballistic (see Fig.~\ref{fig:expansion}). Moreover it appears that the dynamics for repulsive or attractive $U$ in the Fermi-Hubbard model~\eqref{eq:fermi-hubbard} is essentially the same, indicating a possible dynamical symmetry~\cite{Schneider12}. Simulating the dynamics of such an experiment is a daunting task, since no numerical efficient method for a general many-body quantum system is known in two dimensions, and only qualitative features can be reproduced using a Boltzmann equation, whose solutions are shown at the bottom of Fig.~\ref{fig:expansion}.

In Ref.~\cite{Hung2010} transport in the Bose-Hubbard model has been studied from a different perspective by increasing the ratio $U/J$ with different rates to differentiate the time scales involved. A related interesting result is the verification~\cite{Cheneau12} of the Lieb-Robinson bound on the speed of propagation of correlations  reminiscent of the speed of light in relativity~\cite{Lieb1972}. The influence of the superfluid to Mott insulator transition in the presence of a superfluid current has been studied in Refs.~\cite{Altman2005,Polkovnikov2005,Peotta14_SFMI}. The flexibility of varying the interactions of ultracold gases in time or in space can lead to many interesting phenomena. Notable examples are the negative differential conductance in the presence of  spatially inhomogeneous interactions~\cite{ChienNJP13}, where the coupling $U$ in Eqs.~\eqref{eq:bose-hubbard} and \eqref{eq:fermi-hubbard} depends on its position. It was found that energy spectra mismatch leads to a suppression of transport, and this is a generic feature which has also been found in the suppression of energy exchange between atoms with different internal states ~\cite{McKay13}, an effect that is relevant to further cooling down strongly interacting atomic gases.

Snapshots of \textit{in situ} density profiles with single particle resolution grant access to the statistics of fluctuations~\cite{Chien12,Jacqmin2011,Armijo2012}, which has stimulated the study of noise in ultracold gases~\cite{Klawunn2011,Boyang2014}. The fluctuations provide information on the entanglement entropy between different regions of the system. For a bipartite system consisting of two subsystems $A$ and $B$, the entanglement entropy is $s=-\mathrm{Tr}(\rho_{A}\log_{2}\rho_{A})$, where $\mathrm{Tr}$ denotes the trace and $\rho_{A}$ is the reduced density matrix from tracing out the contributions from subsystem $B$. It has been shown that the entanglement entropy $s$ of noninteracting fermions has a compact expression in terms of the semiclassical full-counting statistics~\cite{Klich2009,Klich2009a,Chien14}, and the variation of the entanglement entropy is directly related to the transmission coefficient of the junction connecting the reservoirs. The entanglement entropy is an important quantity that, e.g., constrains the performances of state of the art numerical methods, such as the Density Matrix Renormalization Group widely used to simulate one-dimensional interacting ultracold gases~\cite{Schollwock11,Kollath05,Enss12,Peotta14_SFMI}.

The interplay between disorder and interaction is another topic which is very difficult to analyze using conventional methods due to many material-dependent mechanisms competing simultaneously. The ability of introducing disorder in a controlled way allows experimental verifications of theoretical predictions. Whereas the Anderson localization~\cite{Anderson58,Abrahams10} showing a transition from a conducting to an insulating state of noninteracting particles in disordered potentials has been demonstrated experimentally using ultracold gases~\cite{Billy08,Roati08,Kondov11}, it has been predicted that interacting bosons in a disordered lattice may form interesting phases such as the Bose glass~\cite{Fisher1989} or a many-body localized phase~\cite{Huse2014}. Using ultracold gases it has been possible to provide experimental evidence supporting the existence of both phases~\cite{Fallani07,Schreiber15}.
Further experiments on the transport properties of these phases may shed light on the nature of the superfluid-insulator transition as the disorder is gradually increased~\cite{Gantmakher2010}.

One particularly exciting opportunity offered by ultracold gases is the study of the dynamics of closed, unitarily evolving, and strongly interacting quantum systems on time scales that are readily accessible experimentally (see Table~\ref{Tab:comparison}). Quite often the focus is on the steady state and the fact that a true steady state forms is usually taken for granted. Investigations in ultracold gases reveal that the transient dynamics itself already exhibit rich physics. For example, a steady state may take an extremely long time to emerge in certain setups,  an effect observed in the so-called \qql quantum Newton's cradle\qqr~\cite{Kinoshita2006}. In this important experiment arguably the simplest possible quantum many-body system has been realized, namely a gas of bosons moving in one dimension and interacting by means of contact interactions. The exact Bethe Ansatz solution for the ground state of the model has been provided by Lieb and Liniger~\cite{Lieb:1963a,Lieb:1963b}, but it is still challenging to calculate the time evolution of simple quantities such as the density profile. In Ref.~\cite{Kinoshita2006} the collision of two initially well separated clouds in a harmonic potential is considered and, surprisingly, it is found that the two clouds do not  readily merge into a single one after few oscillations as it happens in the case of a three dimensional condensate. Rather, the clouds keep bouncing from each other many times before exhibiting relaxation, and similar behavior has been demonstrated in interacting fermions \cite{Sommer11}. It is argued that ultimately the system relaxes due to the presence of the harmonic confining potential, whereas in the exactly integrable Lieb-Liniger model the steady state is argued to be distinct from the thermal equilibrium one~\cite{Rigol:2013}. This has stimulated studies of themalization in closed quantum systems~\cite{Trotzky2012}.
Other interesting examples of dynamics in closed quantum systems are the motion of an impurity in a gas of majority atoms of a different kind~\cite{Palzer2009,Fukuhara2013,Catani2012} with interesting connections to the polaron problem in condensed matter physics, and the dynamics of shock waves in quantum gases~\cite{Dutton2001,Hoefer2006,Meppelink2009,Joseph2011}, which is again made possible by the available techniques for measuring the density profile in real time. The propagation of nonlinear waves in various kinds of atomic clouds has been considered in a number of theoretical works~\cite{Bettelheim2012,Kulkarni2012,Lowman2013,Protopopov2013,Peotta14_ShockWave,Peotta14_SFMI},  
and there are studies contemplating various flow patterns, such as dynamical stripe phases induced by fermionic statistics~\cite{Beria13,Yamamoto13,Chern14}.

\section{Conclusion and outlook}
In summary, transport phenomena in cold atoms closely resemble those in conventional solid state systems while exhibiting subtle but fundamental differences as summarized in Table~\ref{Tab:comparison}. In addition to serving as powerful quantum simulators for demonstrating key mechanisms in complex quantum many-body systems, cold-atom systems are capable of probing novel transport phenomena that are considered difficult to access in the solid state. Moreover, cold atoms open the door for systematic studies of quantum transport in isolated quantum systems, where the system can be driven out of equilibrium by conventional ways of coupling to external sources or by ``internal knobs'' such as variable local interactions, density, geometry, etc. In addition to the mass current, dynamical properties such as spatial or momentum distributions can be measured to unravel mechanisms significant in isolated quantum systems~\cite{ChienNJP13,Chien12,Chien14}. While many transport quantities in isolated systems agree with their counterparts in conventional open systems in the thermodynamic limit, in some cases the resemblance turns out to be utterly superficial, for instance the underlying mechanisms that give rise to the quasi-steady state current illustrated in Figure~\ref{fig:QSSC} and the steady state current in the Landauer formalism can be totally different~\cite{Chien14}. 

We also mention that rapid developments of cold-atom systems have initiated a thriving field called ``atomtronics''~\cite{Micheli2004,Albiez2005,Seaman2007,Pepino2009,LeBlanc2011,Lee2013} with the goal to complement and simulate electronic materials, devices, and circuits using ultracold atoms in engineered potentials. Indeed, many interesting applications are expected from this field with advances in the scalability of atom-chips~\cite{Folman2002,Reichel2002,Reichel11}, where patterned circuits on a substrate generate desired magnetic fields to trap and manipulate cold atoms. Optimized designs could potentially lead to portable atomtronic devices of the size of a liter \cite{Rushton14}. The relative ease of loading noninteracting particles \cite{Schneider12}, superfluids \cite{Wright13}, or other many-body phases \cite{Greiner02,Tarruell12,Jo12} in engineered potentials gives atomtronic devices a broader range of applications. As shown in Table~\ref{Tab:comparison}, the tunneling time of cold atoms in artificial lattices can be extremely slow when compared to that of electrons in metals, so it is unlikely atomtronic devices can compete in speed with solid state devices. Instead, the prominent transient behavior and tunability of various parameters could allow atomtronic devices to be used as diagnosis tools for electronics. Moreover, due to the finite holding time of cold atoms, atomtronic devices may be more suitable for single-shot or disposable applications instead of operating in cycles, as commonly seen in solid state devices.  

Transport in conventional electronics has been successful in its applications partly due to a sound understanding of its underlying mechanisms \cite{AshcroftBook,MahanBook,DiVentra_book,Nazarov_book}. In the same vein, the rapid growth in atomtronics and cold-atom technology is poised to bring great opportunities for broadening our understanding of isolated quantum systems and resolving fundamental issues in nonequilibrium physics. Last but no least, novel concepts developed with cold-atom systems may further lead to a paradigm shift in designing and utilizing future quantum devices. 

SP and MD acknowledge support from DOE under Grant No. DE-FG02-05ER46204.
SP acknowledges support from the Academy of Finland through its Centres of Excellence Programme (2012-2017) and under Project No. 251748.

\bibliographystyle{apsrev4-1}
\bibliography{reference}

\end{document}